\documentclass{article}
\usepackage{amsmath,graphicx}\usepackage{amsfonts} 
\usepackage[preprint]{spconf}

\usepackage{color} 
\usepackage[acronym]{glossaries}
\usepackage{amsfonts}

\usepackage{caption}
\usepackage{subcaption}
\usepackage{booktabs}
\usepackage{graphics}
\usepackage{multirow}
\usepackage{url}
\usepackage{enumitem}
\usepackage{flushend}

\title{Resource-constrained stereo singing voice cancellation}
\name{ Clara Borrelli \quad James Rae \quad Dogac Basaran  \quad Matt McVicar \qquad Mehrez Souden \quad Matthias Mauch}

\address{Apple}
\makeglossaries
\newacronym{mss}{MSS}{Music Source Separation}
\newacronym{svc}{SVC}{Singing Voice Cancellation}
\newacronym{stft}{STFT}{Short-Time Fourier Transform}
\newacronym{sisdr}{SI-SDR}{Scale Invariant Source-to-Distortion Ratio}
\newacronym{sir}{SIR}{Source-to-Interference Ratio}
\newacronym{sdr}{SDR}{Source-to-Distortion Ratio}
\newacronym{sar}{SAR}{Source-to-Artifacts Ratio}
\newacronym{mushra}{MUSHRA}{Multiple Stimuli with Hidden Reference and Anchor}
\newacronym{tcn}{TCN}{Temporal Convolutional Network}

\begin{document}
\copyrightnotice{\copyright \begin{tabular}[t]{@{}l@{}}2024 IEEE. Personal use of this material is permitted. Permission from IEEE must be obtained for all other uses,  in any current or \\ future media, including reprinting/republishing this material for advertising or promotional purposes, creating new collective works, \\ for resale or redistribution to servers or lists, or reuse of any copyrighted component of this work in other works.\end{tabular}}

%
\maketitle
\begin{abstract}
We study the problem of stereo singing voice cancellation, a subtask of music source separation, whose goal is to estimate an instrumental background from a stereo mix.
We explore how to achieve performance similar to large state-of-the-art source separation networks starting from a small, efficient model for real-time speech separation.
Such a model is useful when memory and compute are limited and singing voice processing has to run with limited look-ahead. 
In practice, this is realised by adapting an existing mono model to handle stereo input. Improvements in quality are obtained by tuning model parameters and expanding the training set.
Moreover, we  highlight the benefits a stereo model brings by introducing a new metric which detects attenuation inconsistencies between channels.
Our approach is evaluated using objective offline metrics and a large-scale MUSHRA trial, confirming the effectiveness of our techniques in stringent listening tests.

\end{abstract}

\section{Introduction}
\label{sec:intro}

In this work we present a study of optimisation and evaluation of a \gls{svc} system. 
We analyse the performance of such as system in limited computational resource scenarios and understand which factors most affect output quality. \gls{svc} consists of removing the singing voice from the instrumental background in a fully mixed song, and can be interpreted as a special case of \gls{mss}. 
The main difference lies in the fact that \gls{mss} aims to retrieve one separate stem for each source present in the mix, while \gls{svc} considers only instrumental accompaniment as desired source. 

Recent advances in deep learning have led to significant performance improvements in both \gls{svc} and \gls{mss}, leading to much improved objective metrics.
\gls{mss} algorithms typically use a spectral \cite{vardhana2018convolutional} or waveform \cite{defossez2019music} representation. In the first case, both input and output are expressed in the time-frequency domain and popular solutions adopt a U-Net architecture \cite{vardhana2018convolutional, hennequin2020spleeter}, combined with band-split RNN \cite{luo2023bandsplitrnn}, conditioning mechanism \cite{choi2021lasaft} or multi-dilated convolutions \cite{takahashi2020d3net}.
Inspired by Wave-Net \cite{stoller2018wave}, waveform-based \gls{mss} methods tackle the problem end-to-end. An example is Demucs \cite{defossez2019music}, which uses 1-D convolutional encoder and decoders with a bidirectional LSTM. Later versions of Demucs include a spectral branch \cite{defossez2021hybrid} and transformer layers \cite{rouard2023hybrid}.
Similar approaches have been adopted for Singing Voice Separation \cite{choi2020investigating, ni2022fc}.
All of these methods produce high quality results but are often designed to be run offline and in a non-causal setup, i.e., having access to the complete audio input, and no constraints on memory or run-time.
These methods are not suitable for challenging yet realistic scenarios, where separation happens on low-memory edge devices, under streaming conditions and with hard constraints on processing time. 
Efficient solutions have been proposed to tackle speech source separation under these circumstances. Conv-TasNet \cite{luo2019conv} is an end-to-end masking-based architecture which employs 1-D convolutional encoder and decoder, a \gls{tcn}-based separator and depth-wise convolutions to reduce network size. This architecture has also been adapted and expanded to address \gls{mss} \cite{hu2022hierarchic,samuel2020meta}.

In this work Conv-TasNet is adapted and optimised for \gls{svc} such that the output is of comparable quality to more resource-hungry networks. Such architecture guarantees a low-memory footprint thanks to a limited number of parameters, and is able to operate in real-time. We train the model on a large dataset, to achieve high quality output and we show how training dataset size and quality affect the model's output quality.
The main contributions in this work are:
\begin{itemize}[leftmargin=*]
    \item To match a real-world setup, we modify Conv-TasNet to take advantage of stereo input and to produce stereo output. This design improves vocal attenuation consistency across stereo channels, and we propose a new stereo metric to verify this improvement.
    \item We propose a stereo separation asymmetry metric able to measure stereo artifacts and show that the proposed stereo architecture helps to prevent them.
    \item To validate the proposed approach, we conduct experiments to evaluate objectively and subjectively using a MUSHRA-style test.
\end{itemize}
We show that a relatively small and specialised model with appropriate training can reach outstanding high quality comparable to much larger models.

\label{sec:problem_formulation}
\label{subsec:network_architecture}
\begin{figure*}[htb]
\centering
\begin{subfigure}[b]{.84\textwidth}
         \centering
         \includegraphics[width=\textwidth]{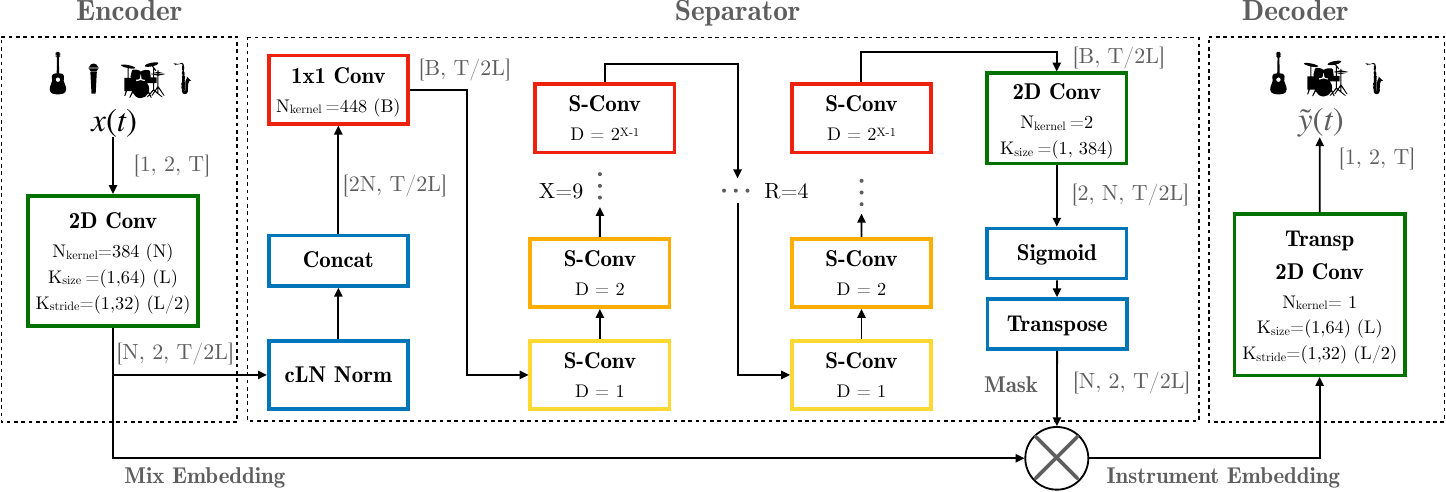}
         \caption{Network Architecture}
         \label{fig:network_architecture}
     \end{subfigure}\hspace{2mm}
     \begin{subfigure}[b]{.132\textwidth}
         \centering
         \includegraphics[width=\textwidth]{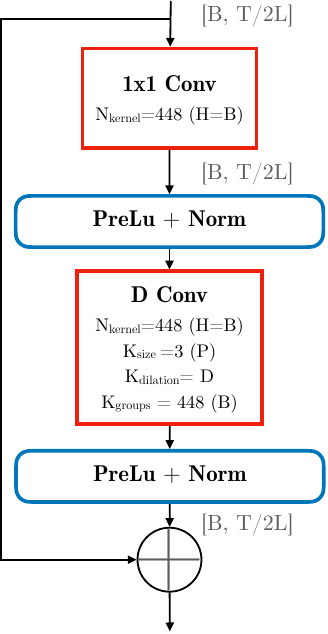}
         \caption{S-Conv Block}
         \label{fig: sconv}
     \end{subfigure}
\caption{Vox-TasNet architecture and S-Conv block in detail. Letters in parenthesis follow the original notation used in \cite{luo2019conv}. In square brackets we report input dimensionality as [Ch, W, H] throughout the network.}
\label{fig:network}
\end{figure*}
\section{Method}
A sampled stereo audio signal $x(t) \in \mathbb{R}^{2}$ with $t \in [0, T)$, which corresponds to a fully mixed track, can be modelled as a linear sum of an instrumental or accompaniment stem $y(t)$ and a vocal stem $v(t)$, i.e.,
The goal of our system is to estimate $y(t)$ in real-time and with low memory requirements.
A real-time scenario implies that the network output should be produced with low-latency.
This forces the model to be almost causal, i.e., the output in a specific time instant depends largely on past input samples and possibly a small portion of future samples, i.e., look-ahead, which can be buffered. 
To be able to satisfy low memory footprint requirements, it is necessary to limit the number of network parameters.
\subsection{Network Architecture}
We propose Vox-TasNet, an adaptation of Conv-TasNet \cite{luo2019conv} to the task of \gls{svc} in a stereo setup, hence able to jointly estimate both left and right channels.
Similar to Conv-TasNet, the network follows a masking-based approach and is composed of three modules, as shown in Fig.~\ref{fig:network}.
The Encoder block reduces time resolution of the raw audio waveform using a two-dimensional convolutional (2D Conv) layer. This creates separate embeddings for the left and right channels of the mix whilst benefiting from the information in both.
These embeddings are stacked and fed into the Separator block. The Separator block consists of multiple stacks of separable-depthwise convolutional (S-Conv) layers with increasing dilation and estimates masks for the left and right channel representations. Only the S-Conv layers in the first group are non-causal, making the Separator block almost entirely causal. This allows the network to look ahead for a small time interval. 
Masking is applied via element-wise multiplication and the resulting masked representations are fed into the Decoder (Transp 2D Conv), which outputs stereo audio accompaniment.
Aside from the stereo setup, there are two main differences between Vox-TasNet and Conv-TasNet. First, the Separator's skip connections are removed to reduce memory footprint, as originally proposed in \cite{pandey2019TCNN}. Second, we increase the number of S-Conv layers in each group but dimensionality is fixed inside each S-Conv layer, rather than expanded and squeezed. This choice allows the separator to learn longer time dependencies, without increasing the number of parameters.
\subsection{Stereo separation asymmetry metric}
\label{subsec:experimentalstereometric}
During informal listening experiments, we noticed that single-channel model produced \gls{svc} results which were audibly inconsistent across channels in terms of loudness and attenuation. On the other side, a stereo-native architecture is able to prevent these artifacts and exploit cross-channel information as well. This lead us to devise a stereo separation asymmetry metric which attempted to measure this effect.

Let's consider a source separation metric, i.e.,  $\text{SI-SDR}$. It can be formulated in a frame-wise setup, i.e., comparing short frames of prediction and ground-truth signal.
Let us define $(\text{SI-SDR}_L(n)$, $\text{SI-SDR}_R(n))$ as the frame-wise metric computed for $N$ windows of length $W$ and hop size $H$ on left channel and right channel respectively, with $n$ corresponding to window index. 
We define $\Delta_{\text{SI-SDR}}(n) = | \text{SI-SDR}_L(n) - \text{SI-SDR}_R(n)|$ as the absolute value of difference between left and right frame-wise metric values.
We use as stereo metric $\text{SSA}_{\text{SI-SDR}} = \frac{1}{N}\sum_n \Delta_{\text{SI-SDR}}(n)$, as the average over time of $\Delta_{\text{SI-SDR}}(n)$ . Larger values of $\text{SSA}_{\text{M}}$ corresponds to large difference between left and right channel's separation metric distance over time, hence quality inconsistency. 
Note that $ \Delta_{\text{SI-SDR}}(n)$ and $\text{SSA}_{\text{SI-SDR}}$ can be defined for any other dB-based source-separation metric (i.e., SIR or BSS-Eval metrics).

\section{Experimental Setup}
\label{sec:experiment}

To train and validate our system, we consider four different datasets. 
Three datasets are internal and  we will refer to them with $\mathcal{A}$, $\mathcal{B}$ and $\mathcal{C}$. 
For benchmarking against state-of-the-art models, we use the publicly available MUSDB \cite{musdb} dataset. In order to be compatible with the task of \gls{svc}, the instrumental accompaniment track is obtained by linearly summing the \textit{bass}, \textit{drums} and \textit{other} stems.
Dataset specifications are reported in Table~\ref{tab:datasets}.
For all internal datasets, we apply a pre-processing step to filter out tracks for which the vocal stem is silent for more than 50\% of the song. 
Note from Table~\ref{tab:datasets} that $\mathcal{C}$ is larger than $\mathcal{A}$ and $\mathcal{B}$, but not musically curated by experts hence potentially contains noisier samples.  

\begin{table}[]
\centering
\resizebox{0.85\columnwidth}{!}{%
\begin{tabular}{llll}
\hline
\textbf{Dataset} &  \textbf{Num Tracks (Train/Test)}& \textbf{Curated} & \textbf{Use} \\ \hline
$\mathcal{A}$    &  3,290         & Yes              & Train        \\
$\mathcal{B}$    & 3,797 (3,375/422) & Yes              & Train/Test         \\
$\mathcal{C}$    & 13,625    & No               & Train        \\
MUSDB            &  150 (100/50) & Yes              & Train/Test   \\ \hline
\end{tabular}
}
\caption{Datasets specifications}
\label{tab:datasets}
\end{table}
To have comparable experiments, we use a fixed set of training parameters in all of our experiments. 
Sampling rate is fixed at $44,100$~Hz, and each training sample is composed of $4$~s of stereo audio, randomly sampled from each song at training time. The batch size is equal to 6 and each model is trained for 500 epochs.
We use the Adam optimizer \cite{kingma2014adam} with an initial learning rate 0.0001, scheduled with decay parameter equal to 0.99.
The loss function is a weighted average of L1 distance in time domain and multi-resolution spectral L1 distance using window lengths equal to $0.01$~s, $0.02$~s and $0.09$~s and hop-size equal to half of window lengths. Time and spectral losses are weighted with weights $0.875$ and $0.125$ respectively.

As a strong baseline, we use HybridDemucs \cite{defossez2021hybrid}, a large music source separation model without real-time or memory constraints.
We aim to match the quality obtained with HybridDemucs, while keeping as reference a resource-constrained scenario.
HybridDemucs is trained on the augmented MUSDB dataset and extracts four separate stems from the original mix: drums, bass, vocal and others. 
We obtain the accompaniment mix as the sum of \textit{drums}, \textit{bass} and \textit{others} stems.
Additionally, we consider a mono version of Vox-TasNet, namely MonoVox-TasNet. It has a similar architecture to Vox-TasNet, as shown in Fig~\ref{fig:network}, but 2-D convolutions are replaced with 1-D convolutions. A stereo output is obtained by separately processing left and right channel and concatenating the results.

To evaluate the proposed system, we take advantage of both objective and subjective evaluation.
For objective evaluation, we use \gls{sisdr} \cite{le2019sdr} computed between the ground-truth accompaniment and the estimated accompaniment. Moreover, we use the separation symmetry metric presented in Section~\ref{subsec:experimentalstereometric}, computed using \gls{sisdr} with $W=1.5$~s and $H=0.75$~s, to evaluate out stereo vs mono model. 
As objective metrics do not always correlate strongly with human perception of audio quality \cite{cano2016evaluation}, we report subjective evaluation results obtained via a \gls{mushra}-style test \cite{series2014method}. For each MUSHRA trial, participants hear a 10-second clip of a mix and reference (i.e., all stems except the vocals). Using the reference as a benchmark, participants use a 0-100 quality scale to rate the output obtained with HybridDemucs, Vox-TasNet, and MonoVox-TasNet trained on $\mathcal{A}$ and $\mathcal{B}$ for 1500 epochs. Also embedded in the test set was the hidden reference (expected to be rated as excellent), as well as a lower anchor (expected to be rated as poor). As a lower anchor we use the original Conv-TasNet architecture without skip connections in the Separator, and trained on $\mathcal{A}$ only. Each participate completed MUSHRA trials for four songs. We exclude judgments from participants if they (a) rated the hidden reference lower than 90 on more than one trial or (b) rated the lower anchor higher than 90 for more than one trial. We test the robustness of our results by comparing subjective model performance on 50 tracks from the MUSDB test set and 53 tracks from the internal datasets obtaining $\mathcal{I}$. Participation was completed (a) online, (b) while using headphones, and (c) after receiving training and the completion of practice trials.  
As MUSHRA scores are bounded (between 0 and 100), we model 2,960 judgements obtained from 99 participants as proportions (dividing each score by 100) using beta regression with a logit link function \cite{ferrari2004beta} and applying a transformation to prevent 0's or 1's \cite{smithson2006better}. Results are transformed back to the original 0-100 scale for ease of interpretation. Dependencies in the data (e.g., each participant made multiple judgements) led us to use a Bayesian multilevel model with weakly informative priors. All parameters met the \(\hat{R} <  1.1\) acceptance criterion \cite{gelman2015bayesian}, indicating model convergence. 

\section{Results}
\label{sec:results}
\subsection{Objective Evaluation}
\label{subsec:objective_evaluation}

\begin{figure}[htb]
\centering
\begin{subfigure}[b]{\columnwidth}
    \centering
    \includegraphics[width=\columnwidth]{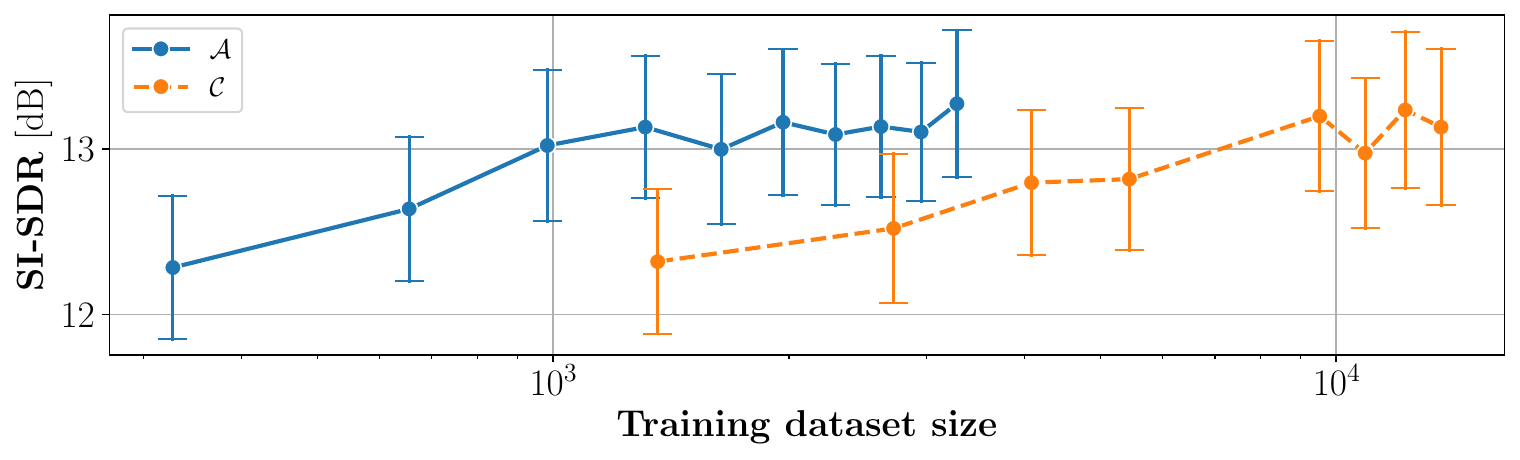}
    \caption{}
    \label{fig:partitions_musdb}
\end{subfigure}
\begin{subfigure}[b]{\columnwidth}
    \centering
    \includegraphics[width=\columnwidth]{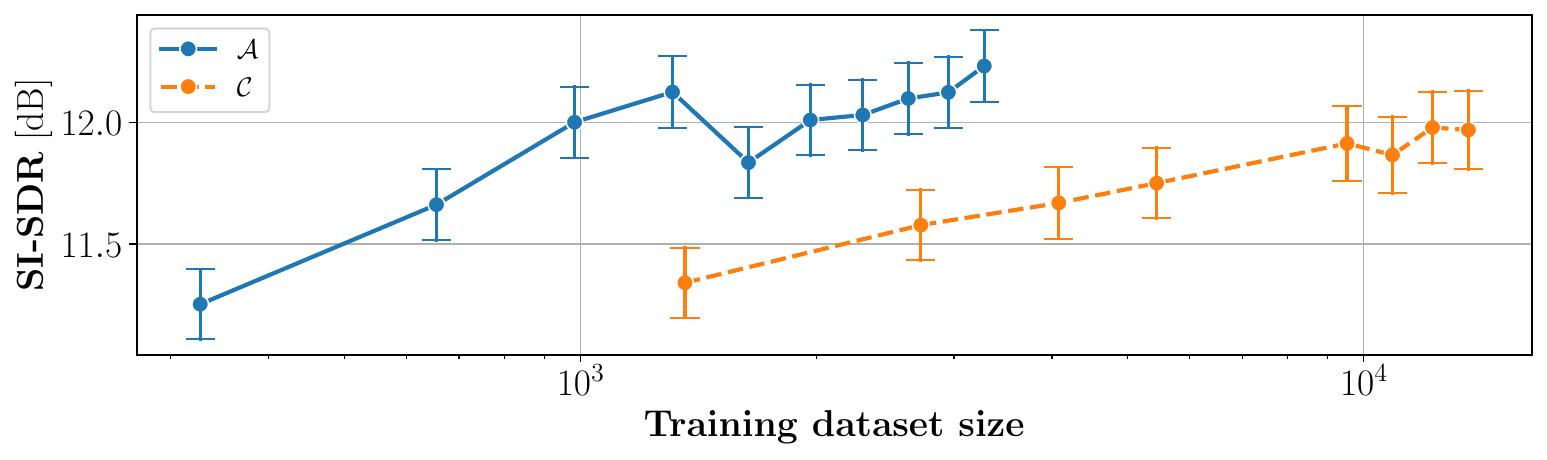}
    \caption{}
    \label{fig:partitions_verm}
\end{subfigure}
\caption{SI-SDR mean and standard error for Vox-TasNet obtained training on different partitions of the two training datasets tested on MUSDB (a) and on $\mathcal{B}$ (b).}
\label{fig:dataset_size}
\end{figure}

We first focus on the analysis of objective metrics on the two evaluation datasets, MUSDB and $\mathcal{B}$. 
In Table~\ref{tab:results_baselines} we show the results for two versions of Vox-TasNet compared to the HybridDemucs baseline, together with correspondent latency and memory footprint.
Both Vox-TasNet versions and the baseline perform differently on the two test datasets. Cross-dataset testing more strongly affects the baseline evaluated on $\mathcal{B}$ when compared to Vox-TasNet trained on $\mathcal{A}$.
Considering \gls{sisdr} values together with the required resources, it is evident how Vox-TasNet trained on $\mathcal{A}$ is able to reach quality comparable to the baseline with less than 10\% network parameters and small look-ahead.
This analysis supports the importance of training dataset in achieving high-quality results while meeting the limited resource constraints.
\begin{table}[]
\centering
\resizebox{\columnwidth}{!}{%
\begin{tabular}{ccccccc}
\hline
\textbf{Model} & \textbf{\begin{tabular}[c]{@{}c@{}}Training \\ dataset\end{tabular}} & \textbf{\begin{tabular}[c]{@{}c@{}}Test \\ dataset\end{tabular}} & \textbf{SI-SDR (dB)}$\uparrow$ & \textbf{\begin{tabular}[c]{@{}c@{}}Receptive\\ field (s)\end{tabular}} & \textbf{\begin{tabular}[c]{@{}c@{}}Look-\\ ahead(s)\end{tabular}} & \textbf{\begin{tabular}[c]{@{}c@{}}Num \\ param (M)\end{tabular}} \\ \hline
\multirow{2}{*}{{\begin{tabular}[c]{@{}c@{}}Hybrid\\ Demucs\end{tabular}}} & \multirow{2}{*}{MUSDB} & MUSDB & $16.34\pm3.03$ & \multirow{2}{*}{5.00} & \multirow{2}{*}{2.5} & \multirow{2}{*}{80} \\
 &  & $\mathcal{B}$ & $14.02\pm3.44$ &  &  &  \\ \hline
\multirow{4}{*}{Vox-TasNet} & \multirow{2}{*}{MUSDB} & MUSDB & $9.90\pm3.22$ & \multirow{4}{*}{1.86} & \multirow{4}{*}{0.37} & \multirow{4}{*}{7.5} \\
 &  & $\mathcal{B}$ & $8.59\pm3.09$ &  &  &  \\
 & \multirow{2}{*}{$\mathcal{A}$} & MUSDB & $13.27\pm3.13$ &  &  &  \\
 &  & $\mathcal{B}$ & $12.23\pm3.28$ &  &  &  \\ \hline
\end{tabular}
}
\caption{Mean and standard deviations of SI-SDR training Vox-TasNet on $\mathcal{A}$ and MUSDB compared to the baseline, together with corresponding latency and memory impact.}
\label{tab:results_baselines}
\end{table}
The second experiment aims at understanding how training dataset size and quality impact model performance. In this case, both $\mathcal{A}$ and $\mathcal{C}$ are partitioned into subsets of increasing size. In Fig~\ref{fig:dataset_size} we report \gls{sisdr} statistics obtained over the number of tracks of training partitions. As expected, in both cases larger training dataset leads to more effective voice cancelling. Comparing the two training datasets at similar dataset size, training on $\mathcal{A}$ always leads to higher \gls{sisdr} and the best result is obtained with the largest $\mathcal{A}$ partition, despite $\mathcal{C}$ being much larger. This result highlights the importance of training data quality, since $\mathcal{C}$ is not musically curated, unlike $\mathcal{A}$.
We also trained the model combining $\mathcal{A}$ and $\mathcal{C}$ datasets with different proportion but no significant improvements were achieved.
The final experiment aims at analysing the differences between the proposed Vox-TasNet stereo model and its mono version, MonoVox-TasNet. Both models have been trained on $\mathcal{A}$. In the first column of Table~\ref{tab:result_monostereo} we report \gls{sisdr} computed separately on left and right channel, indicated as $\text{SI-SDR}^\text{mono}$. We verify that the difference between \gls{sisdr} distributions for mono and stereo model on each evaluation dataset are not statistically significant, hence overall quality is not affected by stereo architecture. In the second column we report values for the stereo metric $\text{SSA}_\text{SI-SDR}$ proposed in Sec~\ref{subsec:experimentalstereometric}. A lower value indicates lower stereo artifacts and higher symmetry between left and right channel vocal attenuation. Results highlight that stereo architecture greatly improves output quality in terms of multichannel artifacts on both test datasets.
\begin{table}[]
\centering
\resizebox{0.85\columnwidth}{!}{%
\begin{tabular}{@{}llll@{}}
\toprule
\textbf{Model} & \textbf{Test Dataset} & \textbf{$\text{SI-SDR}^\text{mono}$ (dB)$\uparrow$} & \textbf{$\text{SSA}_\text{SI-SDR}$ (dB)$\downarrow$} \\ \midrule
\multirow{2}{*}{Vox-TasNet}     & MUSDB         & $12.79\pm3.19$ & $1.10\pm0.45$ \\
                                & $\mathcal{B}$ & $11.64\pm3.19$ & $1.08\pm0.58$ \\
\multirow{2}{*}{MonoVox-TasNet} & MUSDB         & $13.13\pm2.94$ & $1.81\pm0.73$ \\
                                & $\mathcal{B}$ & $11.85\pm3.21$ & $2.30\pm1.34$ \\  \bottomrule
\end{tabular}
}
\caption{Mean and standard deviation of $\text{SI-SDR}^\text{mono}$ and of $\text{SSA}_\text{SI-SDR}$ for Vox-TasNet compared with the mono version trained on $\mathcal{A}$, tested on $\mathcal{B}$ and MUSDB. For $\text{SI-SDR}^\text{mono}$, the higher the better. For $\text{SSA}_\text{SI-SDR}$ the lower the better.}
\label{tab:result_monostereo}
\end{table}
\subsection{Subjective Evaluation}
\begin{figure}[htb]
\centering
\begin{subfigure}[b]{\columnwidth}
    \centering
    \includegraphics[width=\columnwidth]{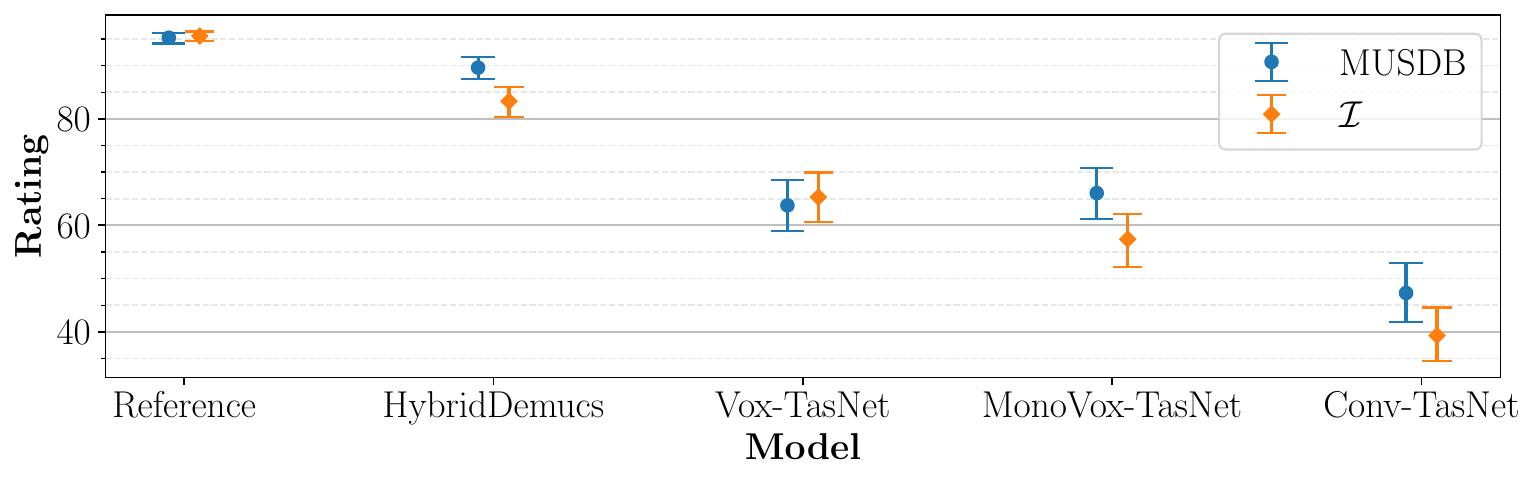}
\end{subfigure}
\caption{Estimated MUSHRA scores for all conditions tested on $\mathcal{I}$ and MUSDB. Intervals are 95\% credible intervals.}
\label{fig:human_evaluation}
\end{figure}

In Fig \ref{fig:human_evaluation} we report MUSHRA scores for all conditions tested $\mathcal{I}$ and MUSDB. As expected, all models had lower scores than the reference audio on each evaluation dataset. Moreover, Vox-TasNet and MonoVox-TasNet had quality scores lower than HybridDemucs, but higher than the original Conv-TasNet architecture. Generally, models were perceived to have higher quality when tested on the MUSDB (vs. $\mathcal{I}$) dataset. Finally, while there was no meaningful difference between the mono and stereo models for the MUSDB dataset, Vox-TasNet had significantly better quality than MonoVox-TasNet on the $\mathcal{I}$ dataset. 

\section{Conclusions}
\label{sec:conclusions}
In this work we presented an efficient stereo \gls{svc} architecture, able to operate in real-time and with low memory requirements. By training the model on a large dataset, we reached performances comparable to larger and non-real-time models. Moreover, we show the benefits of using a stereo architecture through a new stereo separation asymmetry metric which can be formulated for any source separation metric. The results from objective evaluation are validated through a large-scale \gls{mushra} test. We believe this study may help in highlighting the key factors that enable the use of deep learning in real-time music processing. 
\clearpage
\bibliographystyle{IEEEbib}
\bibliography{bibliography.bib}

\end{document}